\documentclass{article}

\PassOptionsToPackage{numbers, compress}{natbib}

\usepackage[final]{neurips_2023_ml4ps}

\usepackage[utf8]{inputenc} 
\usepackage[T1]{fontenc}    
\usepackage{hyperref}       
\usepackage{url}            
\usepackage{booktabs}       
\usepackage{amsfonts}       
\usepackage{nicefrac}       
\usepackage{microtype}      
\usepackage{xcolor}         
\usepackage{graphicx}
\usepackage{amsmath}

\title{Generative Diffusion Models for Lattice Field Theory}

\author{
  Lingxiao Wang \\
    Frankfurt Institute for Advanced Studies\\
    \& Xidian-FIAS International Joint Research Center\\
    Frankfurt am Main,  D-60438, Germany \\
    \texttt{lwang@fias.uni-frankfurt.de}\\
\And
   Gert Aarts \\
    Department of Physics\\
    Swansea University\\
    SA2 8PP, Swansea, United Kingdom\\
    \\
    European Centre for Theoretical Studies in Nuclear Physics and Related Areas (ECT*)\\
    \& Fondazione Bruno Kessler Strada delle Tabarelle\\
    286, 38123 Villazzano (TN), Italy\\
    \texttt{g.aarts@swansea.ac.uk}\\
\And    
    Kai Zhou \\
    Frankfurt Institute for Advanced Studies\\
    Frankfurt am Main,  D-60438, Germany \\
    \texttt{zhou@fias.uni-frankfurt.de}\\
}

\begin{document}

\maketitle

\begin{abstract}
  This study delves into the connection between machine learning and lattice field theory by linking generative diffusion models (DMs) with stochastic quantization, from a stochastic differential equation perspective. We show that DMs can be conceptualized by reversing a stochastic process driven by the Langevin equation, which then produces samples from an initial distribution to approximate the target distribution. In a toy model, we highlight the capability of DMs to learn effective actions. Furthermore, we demonstrate its feasibility to act as a global sampler for generating configurations in the two-dimensional $\phi^4$ quantum lattice field theory.
\end{abstract}

\section{Introduction}

In lattice field theory, physical observables are obtained by approximating path integrals via summing over field configurations, traditionally done through Monte Carlo methods. However, these methods can be computationally expensive. A promising alternative lies in employing generative models, a machine learning framework, to create new configurations following the target physical distribution, thus potentially enhancing the efficiency of Monte Carlo simulations~\cite{Zhou:2023pti,Cranmer:2023xbe}.

Generative models fall into two main categories based on likelihood estimation methods. Implicit Maximum Likelihood Estimation (MLE) employs, for instance, Generative Adversarial Networks (GANs) to generate new configurations through a min-max game, shown effective in lattice simulations~\cite{Zhou:2018ill,Pawlowski:2018qxs}. Explicit MLE uses explicit probability descriptions, e.g., autoregressive models~\cite{Wang:2020hji} and flow-based models~\cite{Albergo:2019eim,Kanwar:2020xzo,Nicoli:2020njz,deHaan:2021erb,Gerdes:2022eve,Chen:2022ytr}, improving efficiency in lattice simulations without needing prepared training data. Despite being plagued by model-collapse and scalability~\cite{Hackett:2021idh,Abbott:2022zsh,Nicoli:2023qsl}, flow-based models have developed rapidly and yielded achievements~\cite{Cranmer:2023xbe}. Recently, Diffusion Models (DMs), a new implicit MLE class but with an explicit probability description, have shown promise in generating high-quality images via stochastic processes~\cite{yang:2022diffusion}, hinting at potential applications also in high-energy physics~\cite{Mikuni:2022xry,Mikuni:2023dvk}.

This work explores the potential of DMs in generating lattice field configurations, and its intrinsic connection with stochastic quantization (SQ)~\cite{Parisi:1980ys,Damgaard:1987rr,Namiki:1992wf} from a stochastic differential equation (SDE) perspective. We demonstrate its efficiency and accuracy of learning effective actions in a toy model, and verify its feasibility of generating configurations in a two-dimensional $\phi^4$ lattice field theory. Our findings suggest that DMs can serve as a significant tool to address computational challenges in lattice simulations, encouraging further explorations in this direction.

\paragraph{Related Work} Langevin dynamics has been utilized in Bayesian learning as a stochastic gradient optimization \cite{welling:2011bayesian}, which introduces stochasticity into the parameter updates, thereby avoiding collapses into local minima. Recent related work introduces stochasticity into the hybrid Monte-Carlo algorithm~\cite{Robnik:2023pgt} and explores the correspondence between the exact renormalizing group (ERG) and DMs based upon the heat equation~\cite{Cotler:2023lem}. From a flow-based model perspective, Ref.~\cite{albergo:2023building} designed a continuous-time normalizing flow with an inferred velocity field from the probability current of a time-dependent density that interpolates between the prior and target densities stochastically.

\section{Stochastic Differential Equation}

In DMs, a denoising model reconstructs original data from its noisy version obtained through a diffusion process. The predetermined process of adding noises, also called the forward process, smoothens the data's probability distribution by introducing noise. The denoising model learns the inverse process—eliminating this noise. Once trained, it generates samples following the data distribution through a reverse diffusion process, starting with random noise and iteratively "cleaning" it until a convergent sample resembling the target data distribution is achieved.

\begin{figure}[htbp!]
\begin{center}
\includegraphics[width=0.8\textwidth]{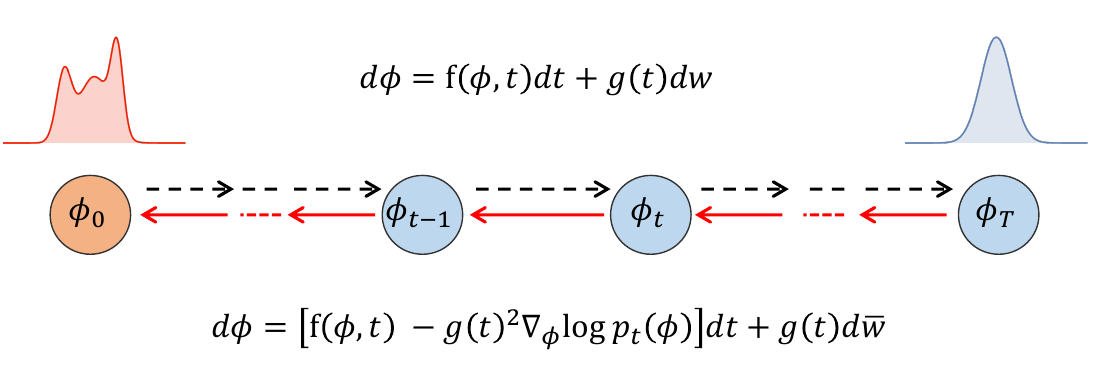}
\caption{A sketch of the forward diffusion process (upper arrows) and the reverse denoising process (bottom arrows). The two stochastic processes are described by two stochastic differential equations. The target distribution is typically unknown but learnt from the initial data.
}
\label{fig:dm}
\end{center}
\end{figure}

The forward process mentioned above can be introduced into the propagation of the field, $\phi_i$, via the following Markov chain, $\phi_i = \phi_{i-1} + f_{i-1} + g_{i-1} \mathbf{z}_{i-1}$, with a total of $N$ steps $(i= 1,\cdots, N)$ and random noise $\mathbf{z}_{i}\sim \mathcal{N}(\mathbf{0}, \mathbf{I})$. Both the random noise $\mathbf{z}_{i}$ and drift force $f_i$ have the same dimensionality as the field $\phi_{i}$. Given an time interval $T \equiv N dt$ and $N\rightarrow\infty$ ($dt\rightarrow 0$), the above forward process converges to its continuous-time limit, which follows an It\^o SDE, $d \phi = \mathbf{f}(\phi, t) dt  + g(t) d \mathbf{w}$, where $t\in[0, T]$, $\mathbf{w}$ is the standard Wiener process, i.e, Brownian motion, $\mathbf{f}(\phi, t)$ is the \textit{drift term},
%
and $g(t)$ is the scalar \textit{diffusion coefficient}. The forward diffusion process $\phi(t)$ can be modeled as the solution of such a generic SDE.

As Figure~\ref{fig:dm} demonstrates, starting from a sample taken from the prior distribution $p_T$ and reversing the above diffusion process enables obtaining a sample from the data distribution $p_0$. Importantly, the reverse process represents a diffusion process evolving backward in time, which is expressed by the following reverse SDE~\cite{anderson:1982reversetime}, $  d\phi = [\mathbf{f}(\phi, t) - g^2(t)\nabla_{\phi}\log p_t(\phi)] dt + g(t) d\bar{\mathbf{w}},$ where the reverse time $t \equiv T-t$, and $p_t(\phi)$ is the probabilistic distribution at time-step $t$, $\bar{\mathbf{w}}$ is a Wiener process in the reverse time direction, and $dt$ represents an infinitesimal negative time step. This reverse SDE can be solved once we know the drift term and diffusion coefficient of the forward SDE, and in particular  $\nabla_{\phi}\log p_t(\phi)$ for each $t\in[0, T]$. 

The reverse SDEs of DMs are mathematically related to Langevin dynamics. For a concise implementation, we choose the \textit{variance expanding} picture of DMs, i.e.\ setting $f(\phi, t) \equiv 0, g(t) \equiv g_\tau$. Its Langevin equation (labeled by a new reverse time $\tau$) now reads,
\begin{equation}
    \frac{d\phi}{d\tau} = g_\tau^2\nabla_\phi \log{p_{\tau}(\phi)} + g_\tau\bar{\eta}(\tau), \label{eq:dmle}
\end{equation}
where the noise term $\langle\bar{\eta}(\tau)\rangle = 0, \langle\bar{\eta}(\tau)\bar{\eta}(\tau')\rangle = 2\bar\alpha\delta(\tau-\tau')$, with $\bar\alpha$ being the diffusion constant. Solving the reverse SDE~\eqref{eq:dmle} to depict denoising is difficult due to the intractable ``time-dependent'' drift term. A U-Net neural network is used to parameterize the score function, ${\mathbf{s}}_\theta(\phi,\tau)$, which estimates the drift term, $-\nabla_\phi \log{p_{\tau}(\phi)}$, in Eq.~\eqref{eq:dmle}. The U-Net accepts time and a trajectory configuration as inputs and outputs the same size as the input. More details about the architecture of the U-Net can be found in Ref.~\cite{Wang:2023exq}.

\subsection{Stochastic Quantization}

In field theory, as an alternative quantization scheme, one can introduce SQ for real actions~\cite{Parisi:1980ys,Damgaard:1987rr}, or complex Langevin dynamics for complex actions~\cite{Parisi:1983mgm,Damgaard:1987rr}. Starting from a generic Euclidean path integral, $Z = \int D\phi\, e^{-S_E}$, SQ introduces a \textit{fictitious} time $\tau$ for the field $\phi$, whose evolution is described by Langevin dynamics,
\begin{equation}
    \frac{\partial \phi(x,\tau)}{\partial \tau} = - \frac{\delta S_E[\phi]}{\delta \phi(x,\tau)} + \eta(x,\tau),
    \label{eq:sq}
\end{equation}
where the noise term  satisfies $\langle \eta(x,\tau) \rangle = 0$, $\langle \eta(x,\tau)\eta(x',\tau') \rangle = 2\alpha\delta(x-x')\delta(\tau - \tau')$, with $\alpha$ being the diffusion constant. In the long-time limit, for real actions
the system reaches an equilibrium state $P_{\text{eq}}(\phi)\propto \exp(-S_E(\phi)/\alpha)$, which follows from properties of the associated Fokker-Planck Hamiltonian~\cite{Damgaard:1987rr}.
For complex actions, there are additional criteria to consider \cite{Aarts:2009uq}.
 
Comparing Eqs.~\eqref{eq:dmle} and \eqref{eq:sq}, one notices the presence of $g^2_\tau$, which rescales both the drift term and the variance of the noise, and is known as a kernel \cite{Damgaard:1987rr}.
Its effect can be absorbed by rescaling time with $g^2(\tau)$, or equivalently absorbing it in the time step, $ g_\tau^2\Delta\tau$. 
One can then identify the drift term in Eq.~\eqref{eq:dmle} with the gradient of an effective DM action $S_\text{DM}$, using $\nabla_\phi S_\text{DM}(\phi,\tau)\equiv -\nabla_\phi \log{p_{\tau}(\phi)} \approx {\mathbf{s}}_\theta(\phi,\tau)$. In the $\tau\rightarrow T$ limit, the distribution $ p_{\tau=T}(\phi) \rightarrow P[\phi,T] \propto \exp(-S_\text{DM}/\bar{\alpha})$. Upon identifying $\bar\alpha$ and $\alpha$,
%
%
%
this implies that the equilibrium state from a SQ perspective can be obtained by denoising a naive distribution using the DM prescription. Concurrently, sampling from a DM is equivalent to optimizing a stochastic trajectory to approach the equilibrium state in Euclidean quantum field theory, $P_{\text{eq}}[\phi] \propto \exp(-S_E/\alpha)$,
This will be demonstrated in the following Section.

\begin{figure}[hbtp!]
    \centering
    \includegraphics[width = 0.8\textwidth]{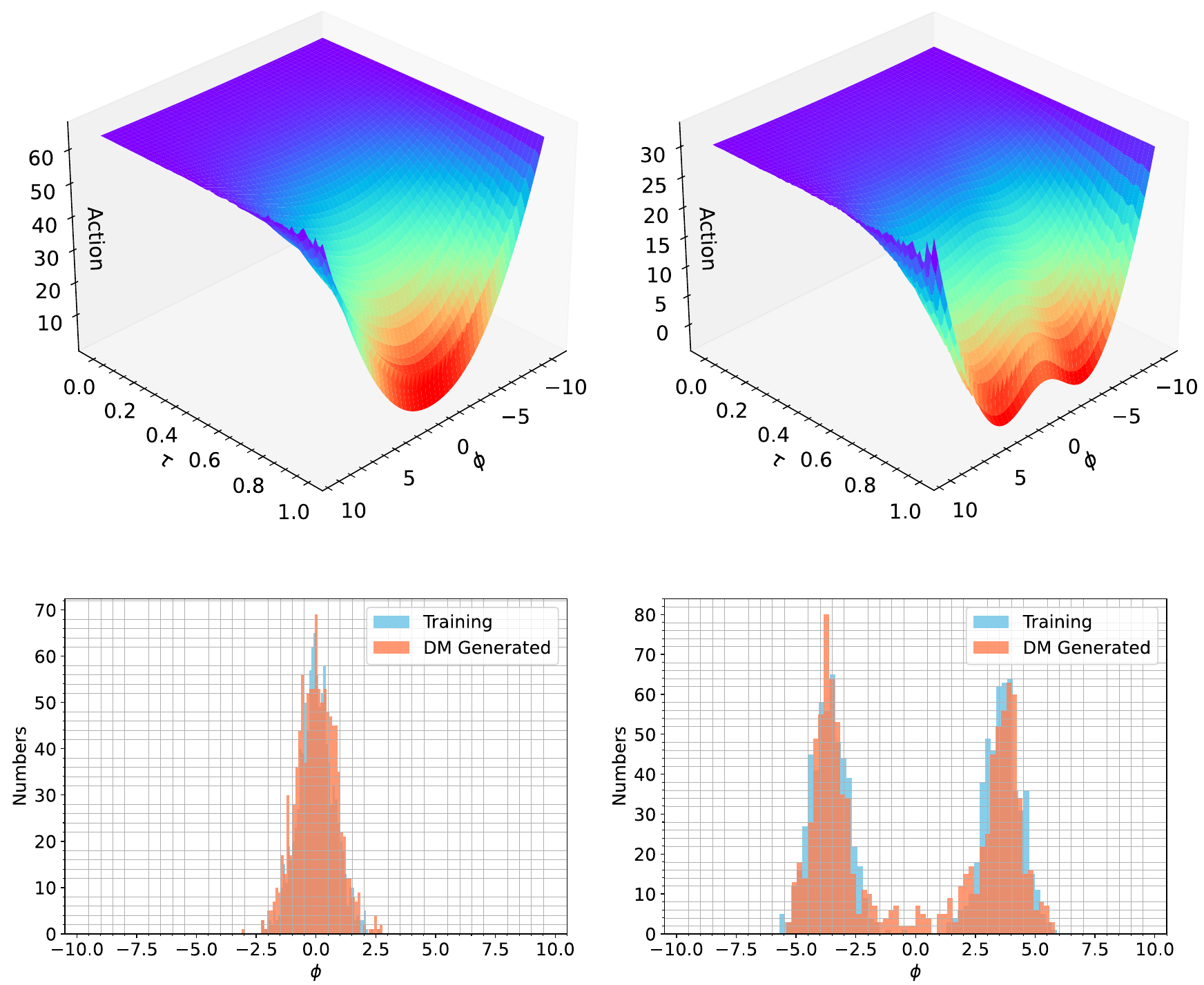}
    \caption{\textbf{(Upper panel) }The flow of the effective action, $S_\text{DM}(\phi,\tau)$, for various values of time $0\leq \tau\leq T=1$ during the stochastic process, learned by the diffusion model as a function of $\phi$ for both the single-well (left column) and double-well (right column) actions, using the relation $\nabla_\phi S_\text{DM}(\phi,\tau) = -\nabla_\phi\log p_\tau(\phi)\approx {\mathbf{s}}_\theta(\phi,\tau)$. \textbf{(Bottom panel)} The first 1024 training samples (blue histograms) and 1024 generated samples (orange histograms) for both the single-well (left column) and double-well (right column) actions.}
    \label{fig:effective_actions}
\end{figure}

\section{Numerical Results}
\label{sec:numerical}


\paragraph{Toy Model} 
To demonstrate the capacity for learning effective DM actions, $S_\text{DM}(\phi,\tau)$,  defined via the relation $\nabla_\phi S_\text{DM}(\phi,\tau) = -\nabla_\phi\log p_\tau(\phi)\approx {\mathbf{s}}_\theta(\phi,\tau)$, 
we introduce an oversimplified $0+0$-dimensional field theory, i.e., a toy model with only one degree of freedom, and the action and drift term,
\begin{equation}
    S(\phi) = \frac{\mu^2}{2}\phi^2 + \frac{g}{4!}\phi^4,
    \qquad\quad
     f(\phi) = -\frac{\partial S(\phi)}{\partial\phi} = - \mu^2 \phi - \frac{g}{3!}\phi^3,
\end{equation}
with parameters $\mu^2$ and $g$. We prepared 5120 configurations as training datasets in two setups: $\mu_1^2 = 1.0, g_1 = 0.4$ (single-well action), and $\mu_2^2 = -1.0, g_2 = 0.4$ (double-well action). A one-to-one neural network with time-embedding is implemented to represent the score function $\mathbf{s}_\theta(\phi,\tau)$.

After 500 epochs of training, the learned effective action $S_\text{DM}(\phi,\tau) = \int^{\phi}\hat{\mathbf{s}}_\theta(\phi',\tau) d\phi'$ is seen to approximate the action $S(\phi)$ in the upper panel of Fig.~\ref{fig:effective_actions}; it approaches the physical action as $\tau$ increases. We have added to $S_\text{DM}$ a constant $\Delta S_0$, which is the difference between $\min[S(\phi)]$ and $\min[S_\text{DM}(\phi,\tau)]$.  Generally, the learned effective actions are accurate approximations in both the single and double-well cases, around $\phi\sim 0$. In the bottom panel of Fig.~\ref{fig:effective_actions}, samples generated from the trained DM are compared with samples from the underlying theory. In this case, we utilized an Apple M2 Pro with 16GB of RAM and PyTorch for model training, achieving a total training time of 35 seconds over 500 epochs.

\paragraph{Scalar Lattice Field Theory}
We consider a real scalar field in $d$ \textit{Euclidean} dimensions with the dimensionless action, 
\begin{equation}
S_E = \sum_x \left[  -2\kappa \sum_{\nu = 1}^d\phi(x)\phi(x+\hat{\nu}) + (1-2\lambda)\phi^2(x) + \lambda\phi^4(x) \right],
\label{eq:phi4action}
\end{equation}
where $\kappa$ is the hopping parameter, and $\lambda$ denotes the dimensionless coupling constant describing field interactions. Both parameters are positive, more details can be found in Ref.~\cite{Smit:2002ug}.

\begin{figure}[hbtp!]
    \centering
    \includegraphics[width = 0.5\textwidth]{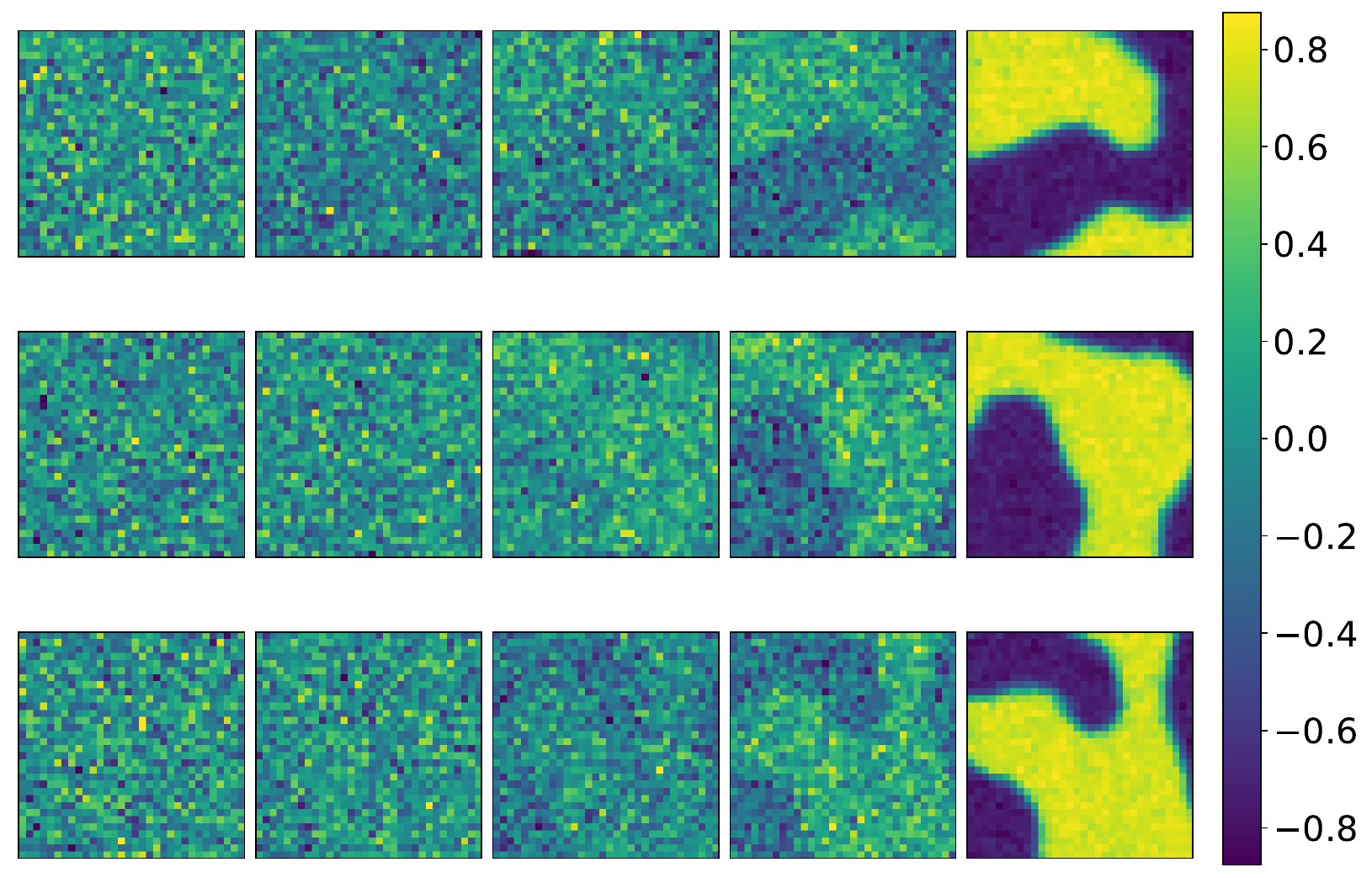}
    \caption{Generation of four independent configurations from a well-trained diffusion model in the broken phase.  Each row in the figure represents a different sample, and each column represents a different time point ($\tau \in [0,0.25,0.5,0.75,T=1]$) during the denoising process.}\label{fig:sample}
    \label{fig:sample_k0.5}
\end{figure}

In the broken phase ($\kappa = 0.5, \lambda = 0.022$), field configurations behave like large clusters. We demonstrate that the clustering behavior of field configurations in a $d=2$ dimensional case, in which it can be successfully captured by the well-trained DM. Fig.~\ref{fig:sample} visualizes the denoising process. The first column represents noise samples randomly drawn from the prior normal distribution, while the fifth column represents the generated samples obtained by denoising. Training set-ups and more quantitative evaluations both in the broken phase and symmetric phase can be found in Ref.~\cite{Wang:2023exq}.

\section{Conclusion and Outlook}

In this contribution, a novel method is introduced to generate quantum field configurations using generative diffusion models. The connection with stochastic quantization is highlighted, an approach to quantize field theories based on a stochastic process in a fictitious time direction. In DMs, the drift term is learned from prepared configurations in a forward process, whereas in SQ it is known and derived from the physical action. The approach is demonstrated in a toy model and a two-dimensional scalar $\phi^4$ field theory. Future directions include further exploring the connection between DMs and SQ, training DMs without a training data set, and using DMs to incorporate the effect of fermions in QCD. Additionally, combining DMs with complex Langevin dynamics may be used to generate configurations for theories with a sign problem, such as e.g.\ QCD at nonzero baryon density \cite{Aarts:2015tyj}. 

\begin{ack}
We thank Profs.\ Tetsuo Hatsuda, Shuzhe Shi and Xu-Guang Huang for helpful discussions. 
We thank ECT* and the ExtreMe Matter Institute EMMI at GSI, Darmstadt, for support during the ECT*/EMMI workshop {\em Machine learning for lattice field theory and beyond} in June 2023 during the preparation of this paper.
The work is supported by (i) the BMBF funded KISS consortium (05D23RI1) in the ErUM-Data action plan (K.Z.), (ii) the AI grant of SAMSON AG, Frankfurt (K.Z.\ and L.W.), (iii) Xidian-FIAS International Joint Research Center (L.W), (iv) STFC Consolidated Grant ST/T000813/1 (G.A.). K.Z.\ also thanks the donation of NVIDIA GPUs from NVIDIA Corporation.
\end{ack}

\bibliographystyle{unsrt}
\bibliography{langevin_mlps.bib}

\end{document}